# SOLUTION OF THE EQUATIONS
# OF ELECTROMAGNETIC SELF-CONSISTENCY
# IN A PLASMA

**Evangelos Chaliasos**

365 Thebes Street

GR-12241 Aegaleo

Athens, Greece

*Abstract*

The system of equations of electromagnetic self-consistency in a plasma is analytically solved for the case of a two-component homogeneous plasma in the non-relativistic approximation.

## 1. Introduction

It is shown in this paper that the equations of electromagnetic self-consistency in a (two-component) plasma, found in a previous paper [Ch] as being 14 equations (12 main equations plus 2 auxiliary equationsa) in 12 unknowns, have a solution for a homogeneous non-relativistic plasma, since any two of them are compatible with the remaining ones.

We expose the equations and specialize the problem to the case of a "pressureless" plasma (strictly speaking to one without pressure gradient), in sec. 2.

We work necessarily in the Lorentz gauge, since the equations contain this gauge. Then equation (2), without the pressure term, is solved in sec.3.

We insert the solution found in eq. (1), and this is reduced to a solvable form, because of the equation of continuity, in sec.4, in which section we also introduce some simplifications as being necessary.

In sec.5 we solve this equation (analytically).

**2. Exposition of the equations and specialization of the problem**

The main equations of electromagnetic self-consistency in a (two-component) plasma are [Ch]

$$\frac{c}{4\pi}\frac{\partial^2 A^l}{\partial x_i \partial x^i} = \sum_{+,-}\left(j_{\pm}^{\ l}\right) \quad \text{(Maxwell equations)} \tag{1}$$

$$\frac{1}{c}\left(A_i \frac{\partial n_{\pm}}{\partial x^k} - n_{\pm}\frac{\partial A_k}{\partial x^i}\right)u_{\pm}^{\ i} + \frac{c}{\lambda_{\pm}}\left(\frac{\partial n_{\pm}}{\partial x^k} - n_{\pm}u_{\pm}^{\ i}\frac{\partial u_{\pm k}}{\partial x^i}\right) = 0 \text{ (equation of motion)} \tag{2}$$

(12 equations with 12 unknowns), with the auxiliary equations

$$\frac{\partial A^k}{\partial x^k} = 0 \quad (k = 0, 1, 2, 3) \quad \text{(Lorentz condition)} \tag{3}$$

to Maxwell equations (1), and

$$\frac{\partial j_{\pm}^{\ i}}{\partial x^i} = 0 \quad (i = 0, 1, 2, 3) \quad \text{(continuity equation)} \tag{4}$$

to the equations of motion (2), where $n_{\pm}$ & $u_{\pm}^{\ i}$ (1+3=4 independent quantities) are related to the current $j_{\pm}^{\ i}$ by

$$j_{\pm}^{\ i} = cn_{\pm}u_{\pm}^{\ i}. \tag{5}$$

The two auxiliary equations (3) & (4) are necessary, since we used them in finding the main equations (1) & (2) respectively [Ch]. Thus we have 4+2x4 +1+1=14 equations for 12 unknowns ($A^i$ and $j_{\pm}^{\ i}$ [or $n_{\pm}$ & $u_{\pm}^{\ i}$]). Ordinarily, in order for a system of equations to have a solution, the number of equations must agree with the number of unknowns. Thus it is possible in principle any two of our equations to be either compatible or incompatible with the remaining ones. We will see that the former is the case.

Note that the equation of motion (2) can be written as

$$n_\pm \frac{\partial u_{\pm k}}{\partial x_i} u_\pm^i = \left(\frac{\lambda_\pm}{c^2} A_i u_\pm^i + 1\right) \frac{\partial n_\pm}{\partial x^k} - \frac{\lambda_\pm}{c^2} n_\pm A_{k,i} u_\pm^i, \tag{6}$$

or

$$\frac{du_{\pm k}}{ds} = \frac{1}{n_\pm}\left(\frac{\lambda_\pm}{c^2} A_i u_\pm^i + 1\right)\frac{\partial n_\pm}{\partial x^k} - \frac{\lambda_\pm}{c^2}\frac{dA_k}{ds}. \tag{7}$$

In this form, we may identify the term

$$\frac{1}{n_\pm}\left(\frac{\lambda_\pm}{c^2} A_i u_\pm^i + 1\right)\frac{\partial n_\pm}{\partial x^k} \tag{8}$$

with a pressure term[*] , since in a non-fluid approach (that is without pressure), we have only, as an equation of motion,

$$\frac{du_{\pm k}}{ds} = -\frac{\lambda_\pm}{c^2} A_{k,i} u_\pm^i \tag{9}$$

or

$$\frac{du_{\pm k}}{ds} = -\frac{\lambda_\pm}{c^2}\frac{dA_k}{ds}. \tag{10}$$

This is the case because as we have seen [Ch] that equation giving the Lorentz force, namely

$$mc\frac{du^i}{ds} = \frac{e}{c} F^{ik} u_k , \tag{11}$$

has to be replaced by the equation

$$mc\frac{du^i}{ds} = -\frac{e}{c} A^i_{,k} u^k , \tag{12}$$

or

---

[*] Stricktly speaking, since

$$p^i = \frac{n}{\lambda} c u^i \tag{a}$$

(for the case of a fluid), we have for the force

$$\frac{dp^i}{ds} = \frac{c}{\lambda}\frac{dn}{ds} u^i + \frac{c}{\lambda} n \frac{du^i}{ds}, \tag{b}$$

and then (making use of the equation of motion (2)) finally

$$\frac{dp^i}{ds} = \frac{c}{\lambda} u^k \left(\frac{\partial n}{\partial x^k} u^i + u_k \frac{\partial n}{\partial x^i} + \frac{\lambda}{c^2} A_k \frac{\partial n}{\partial x^i}\right) - \frac{n}{c}\frac{\partial A^i}{\partial x^k} u^k \tag{c}$$

Thus, we see from (c) that, the terms containing the four-gradient of n <u>here</u> have to be identified as pressure forces, while the last term in (c) has to be taken as the <u>external force due to the field.</u> Note that then, if we neglect the pressure forces, we find for the "external force"

$$\vec{f} = -\frac{\rho}{c}\left[\frac{\partial \vec{A}}{\partial t} + (\vec{v} \cdot grad)\vec{A}\right], \tag{d}$$

especially in the non-relativistic approximation.

$$mc\frac{du^i}{ds} = -\frac{e}{c}\frac{dA^i}{ds} \ . \qquad (13)$$

In fact Eqs. (12) and (13) become in the case of a fluid, and in the notation of our problem, exactly

$$\frac{du^i}{ds} = -\frac{\lambda}{c^2} A^i{}_{,k} u^k \ , \qquad (14)$$

or

$$\frac{du^i}{ds} = -\frac{\lambda}{c^2}\frac{dA^i}{ds} \qquad (15)$$

(cf. Eq. (9) or (10)). Effectively Eq. (14) or (15) is physically equivalent to (2) (or (6) or (7)) in the absence of a pressure gradient. This is why we identify the term (8) with a pressure term, and this is why they are the same mathematically, if we equate the term (8) to zero. Equating the pressure term to zero amounts to setting

$$\frac{\lambda_\pm}{c^2} A_i u_\pm{}^i + 1 = 0 \qquad (16)$$

*ad hoc,* or assuming that $n_\pm$ are constant. We will see that the latter is the case.

### 3. The Lorentz gauge and the solution of Eq. (2)

First of all we have to work in the Lorentz gauge, as it has been mentioned above. This is so because the Lorentz gauge is itself a part of the set of equations to be solved (these equations are valid *only* in this gauge). Thus we have

$$div\vec{A} + \frac{\partial \phi}{c\partial t} = 0. \qquad (16')$$

We will work for simplicity in the non-relativistic approximation. But this fact will not affect significantly our results, because we are going to apply them to the case of a laboratory plasma.[*]

---

[*] Nevertheless, there are cases of even laboratory plasmas which are not non-relativistic. For example the electrons in the Ion-Channel Laser, which belong to a laboratory plasma, are relativistic [cf. D.H. Whittum et al., Phys. Rev. Lett., 64 (1990) 2511]. Thanks to the anonymous referee, who pointed it out to me.

Starting now the solution of our pressureless (more strictly: without pressure gradient) problem from Eq. (10), we see that solving it is very simple. The solution immediately can be taken as

$$u_{\pm}^{\alpha} = -\frac{\lambda_{\pm}}{c^2} A^{\alpha} \qquad (17)$$

($\alpha = 1,2,3$), apart from the integration constants, which does not damage the generality, since the integration constants can always be incorporated in the potentials $A^{\alpha}$ because, in the Lorentz gauge (16'), we always have this gauge freedom. Here 3 of the 2x3=6 integration constants (corresponding, say, to the + equations) are incorporated directly in the potentials. The remaining 3 integration constants (corresponding to the - equations) are set to equal zero, which means that we consider motion without "initial velocity".

If we take also the k=0 equation into account, we will have

$$u_{\pm}^{0} = -\frac{\lambda_{\pm}}{c^2}(\phi + C_{\pm}), \qquad (18)$$

with $C_{\pm}$ constants of integration. Then we will take the result

$$A^{\alpha} A_{\alpha} = c^4/\lambda_{\pm}^2 - (\phi + C_{\pm})^2, \qquad (19)$$

from which we derive

$$\frac{c^4}{\lambda_{+}^2} - (\phi + C_{+})^2 = \frac{c^4}{\lambda_{-}^2} - (\phi + C_{-})^2. \qquad (20)$$

This equation results in ö = constant, so that we can set it to equal zero without affecting the equations (1) - (4) in our case, the Lorentz gauge (16') thus reducing simply to the Coulomb gauge

$$div\vec{A} = 0, \ \phi = 0. \qquad (20')$$

In fact, Eqs. (17) and (18) satisfy Eq. (2), as long as the condition (cf. (8))

$$\frac{\lambda_{\pm}}{c^3}\frac{\partial n_{\pm}}{\partial x^i} C_{\pm}^2 = 0 \qquad (21) \text{ is}$$

satisfied. But this is indeed the case, because, as we will see in what follows, the charge densities remain constant.

**4. Reduction of Eq. (1) and the equation of continuity**

After equations (17) plus (18), with the aid of (5), equation (1) gives

$$\lambda_+ n_+ C_+ + \lambda_- n_- C_- = 0 \tag{22}$$

(because $\phi = 0$), and

$$\frac{c}{4\pi} \frac{\partial^2 A^\alpha}{\partial x_i \partial x^i} = -\frac{1}{c}(\lambda_+ n_+ + \lambda_- n_-)A^\alpha. \tag{23}$$

From (22), together with (20) with $\ddot{\phi} = 0$, we can determine $C_\pm$ ($n_\pm$ are constants, as we will see in what follows). Equation (23) can be written as

$$\frac{c}{4\pi}\left(\nabla^2 - \frac{1}{c^2}\frac{\partial^2}{\partial t^2}\right)A^\alpha = \frac{1}{c}(\lambda_+ n_+ + \lambda_- n_-)A^\alpha. \tag{24}$$

Confining ourselves to the non-relativistic approximation as we have said, and writing

$$n_+ + n_- = 0, \tag{25}$$

appropriate for a plasma, we can set

$$-n_- = n_+ \equiv n = \rho, \tag{26}$$

where it is implied that

$$u_\pm^0 = 1 \text{ (and } u_\pm^\alpha = \vec{v}_\pm / c). \tag{27}$$

Then, (17) can be written, by means of (5), as

$$\frac{1}{\lambda_\pm}\vec{j}_\pm = \mp\frac{\rho}{c}\vec{A}, \tag{28}$$

or, simpler,

$$\frac{1}{\lambda_\pm}\vec{v}_\pm = -\frac{1}{c}\vec{A}. \tag{29}$$

And also, (24) can be written simply

$$\frac{c^2}{4\pi}D\vec{A} = (\lambda_+ - \lambda_-)\rho\vec{A}, \tag{30}$$

where D denotes the D'Alembertian operator.[*] Equation (1), from which (30) results, for l=0 becomes merely an identity, since $A^0 = \phi = 0$ (because of the Coulomb gauge used, that is (20')) on the one hand, and on the other hand since $\sum_{+,-}(j_\pm^0) = j_{total}^0 = 0$

(because of the assumption (25) and the non-relativistic approximation used).

---

[*] Defined as $D = \nabla^2 - \dfrac{\partial^2}{c^2 \partial t^2}$.

7The equation of continuity (4) becomes now, if we take into account (5),

$$\frac{\partial}{\partial x^i}\left(n_\pm u_\pm^{\ i}\right)=0, \qquad (31)$$

or, expanding,

$$\frac{\partial n_\pm}{\partial x^i}u_\pm^{\ i}+n_\pm\frac{\partial u_\pm^{\ i}}{\partial x^i}=0, \qquad (32)$$

or (because also of (17) & (18)),

$$\frac{dn_\pm}{ds}-\frac{\lambda_\pm}{c^2}n_\pm\frac{\partial A^\alpha}{\partial x^\alpha}=0, \qquad (33)$$

or simply, because of the Coulomb gauge ($div\vec{A}=0$),

$$dn_\pm/ds=0. \qquad (34)$$

Thus, we arrive at the conclusion that both $n_\pm$ are constant, that is (because of (26))

$$\rho=\text{constant}. \qquad (35)$$

Thus, the necessary and sufficient condition for the absence of the pressure term (8) is that the charge densities, and with them the mass density, should remain constant.

Another way to see this is from the non-vanishing of the coefficient of $\partial n_\pm/\partial x^k$ in (8). If we call it $F_\pm$, this can be written as

$$F_\pm=\left(\frac{1}{c}A_i+\frac{c}{\lambda_\pm}u_{\pm i}\right)u_\pm^{\ i}. \qquad (36)$$

Then, because of (17), we obtain

$$F_\pm=\frac{c}{\lambda_\pm}u_{\pm 0}u_\pm^{\ 0}, \qquad (37)$$

and, because of (18),

$$F_\pm=\left(-\frac{1}{c}C_\pm\right)\left(-\frac{\lambda_\pm}{c^2}C_\pm\right), \qquad (38)$$

or

$$F_\pm=\frac{\lambda_\pm}{c^3}C_\pm^{\ 2} \qquad (39)$$

(cf. (21)), that is indeed $F_\pm\neq 0$.

Note that, because of (27), i.e. for $u_\pm^{\ 0}\cong 1$, we obtain from (18)

$$C_\pm\cong-\frac{c^2}{\lambda_\pm}. \qquad (40)$$

In this approximation, we obtain

$$F_\pm \cong c/\lambda_\pm , \tag{41}$$

that is again $F_\pm \neq 0$.

## 5. The final solution

Equation (30) now becomes

$$D\vec{A} = C\vec{A}, \tag{42}$$

where C is a constant, given by

$$C = \frac{4\pi}{c^2}(\lambda_+ - \lambda_-)\rho. \tag{43}$$

We immediately recognize (42) as the eigenvalue equation for the operator D, with $\vec{A}$ the eigenfunction and C the corresponding eigenvalue. We thus obtain as a solution (eigenfunction), any harmonic function of the form

$$\vec{A} = \vec{a}\exp\{-i(\omega t - \vec{k}\cdot\vec{r})\}, \tag{44}$$

with $\vec{a}$ a constant vector, and ω the (cyclic) frequency and $\vec{k}$ the wave vector. The corresponding eigenvalue is then

$$C = \frac{\omega^2}{c^2} - \vec{k}^2. \tag{45}$$

Of course, it must be noted here that, if

$$k^i = (\omega/c, \vec{k}) \tag{46}$$

is the corresponding wave four-vector, it must not be null, that is the expression

$$k^i k_i = \frac{\omega^2}{c^2} - \vec{k}^2 \tag{47}$$

must not vanish (it must be of course $k^i k_i > 0$, i.e. $k^i$ must be timelike).

It has also to be noted that $\vec{a}$ cannot be taken arbitrarily, but, instead, it has to be orthogonal to the direction $\vec{k}$ of propagation of the wave. In fact, because the Coulomb gauge (20') must be satisfied, the condition

$$\vec{a}\cdot\vec{k} = a_\alpha k^\alpha = 0 \tag{48}$$

(α=1,2,3) must be satisfied as well, that is the wave solution must be transverse.

## 6. Application

The solution found above may help in heating a homogeneous plasma in a plasma reactor until the plasma reaches the high temperatures necessary for nuclear reactions to take place. Up to now the heating is based mainly on the pinch effect. In this way it is possible to adiabatically compress inward the plasma, which collapses towards the axis of the cylindrical plasma column involved. The temperatures so reached are indeed high enough for nuclear reactions to begin, but this mechanism contains in itself a great disadvantage: it lasts only for an extremely short interval of time. Exactly this can be overcome by the method described below.

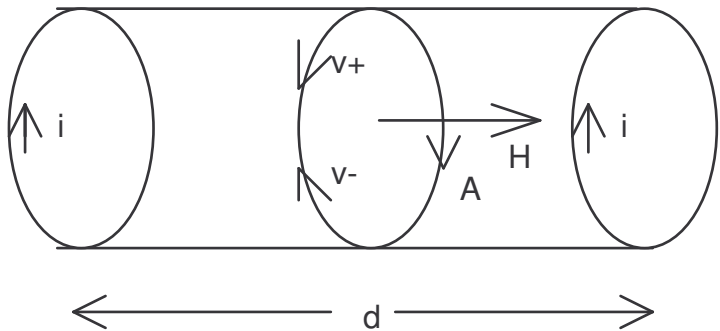

Fig.1

If we are given a cylindrical column of (homogeneous) plasma of length d, as in Fig.1, we place two spirals at its ends, and we supply them with two alternating currents i & i of the same direction, as in the figure, of suitable (cyclic) frequency ω, such that standing waves of $\vec{H}$, with nodes at the ends and antinode in the middle, are created. Thus resonance results in the space of the plasma, between the spirals, concerning the field intensity $\vec{H}$. This intensity results of course from a vector potential $\vec{A}$, which of course will be also in resonance inside the column. Because

$$\vec{H} = curl\vec{A}, \tag{49}$$

and because of the positions of the spirals, $\vec{H}$ and $\vec{A}$ will be as in the figure.

Then, because of (29) and (44), the plasma can be said to be "excited" and two streams of particles (namely ions and electrons) of opposite velocities will be generated, which both will move in circles rotationally oscillating in this way about the axis of the cylinder (see Fig.1), since their velocities $\vec{v}_\pm$ are colinear to $\vec{A}$. Thus, collisions of electrons with ions (protons) will take place continuously, resulting in heating the plasma inside the column.

Equating the right hand side of formulae (43) and (45), we can determine the resonant frequency at which the resonance takes place, by setting $k \equiv 2\pi/\lambda = \pi/d \ (d = \lambda/2)$. The result is

$$\omega = c\sqrt{\frac{\pi^2}{d^2} + \frac{4\pi}{c^2}(\lambda_+ - \lambda_-)\rho}. \tag{50}$$

This frequency falls, for usual dimensions $(\lambda \approx 1m)$, in the limits of radio-waves with micro-waves.

REFERENCES

E. Chaliasos, "The equations of electromagnetic self-consistency in a plasma",*Commun. Theor. Phys.* (Beijing, China) **39** (2003) 711.